\documentclass{PoS}
\usepackage{graphicx}
\usepackage{amsmath}

\title{Magnetic Moments of Negative-Parity Baryons from Lattice QCD}

\ShortTitle{Magnetic Moments of Negative-Parity Baryons}

\author{\speaker{Frank X. Lee}\\
        Physics Department, The George Washington University, Washington, DC, USA\\
        E-mail: \email{fxlee@gwu.edu}}

\author{Andrei Alexandru\\
        Physics Department, The George Washington University, Washington, DC, USA\\
        E-mail: \email{aalexan@gwu.edu}}

\abstract{We report preliminary results for the magnetic moments of negative-parity baryons extracted
from mass shifts in the presence of static external magnetic fields.
The calculations are done on $24^3\times 48$ quenched lattices using standard Wilson actions,
with $\beta$=6.0 and pion mass down to about 520 MeV, and 1000 configurations. 
Reasonable signals for the negative-parity states are observed and the sign 
of their magnetic moments is established. 
The results are compared to model calculations.}

\FullConference{The XXVIII International Symposium on Lattice Filed Theory\\
                June 14-19,2010\\
                Villasimius, Sardinia Italy}

\begin{document}

 \section{Introduction}
Excitation of the nucleon is an interesting problem in nuclear and particle physics.
Negative-parity partners of the baryon octet arise from excitation of one unit 
of orbital angular momentum, and their mass splittings can be traced to spontaneous 
breaking of chiral symmetry of QCD. Given that the mass spectrum of the 
$1/2-$ spin-parity states has been reasonably well established from lattice QCD calculations~\cite{Mahbub10,Bulava10,Mathur05,Sasaki02},
it is instructive to investigate the magnetic moment of the states.
Magnetic moment is the leading-order response of a bound system to a soft external 
magnetic field, and offers a venue to peek into its internal structure and 
the inner workings of QCD. 
Although the magnetic moments of the $1/2+$ baryon octet are well-known both experimentally and 
theoretically, little is known about their $1/2-$ counterparts. 
Experimentally, they can be accessed in photo- and electro-production of 
mesons at intermediate energies, but to date no such measurements have been made.
It would be interesting to see what QCD predicts for these negative-parity states.

 \section{Method}
It is known as the background field method~\cite{mart82,bernard82,rubin95,Lee05,Lee06,Engel07,Aubin09,Detmold10}.
For a particle of spin $S$ in an uniform magnetic field, the interaction energy is 
\begin{equation}
E_\pm=m + \mu B s + \cdots,
\label{energy}
\end{equation}
where the ellipses indicate higher-order terms in the magnetic field, 
and $s=S_z/S$ (for spin-1/2 particles $s=\pm 1$). 
The upper sign means spin-up and the lower sign means spin-down relative 
to the magnetic field.
The magnetic moment is related to the so-called g-factor by $\mu=g {e\over 2m}$. 
So by computing the mass shift $\delta m=E(B)-m$ in the presence of a small 
background magnetic field, one can extract the magnetic moment.

In order to place a magnetic field on the lattice, we construct an 
analogy to the continuum case. The fermion action is modified 
by the minimal coupling prescription 
$D_\mu = \partial_\mu+gG_\mu + q A_\mu $
where $q$ is the charge of the fermion field and $A_\mu$ is the vector 
potential describing the background field. On the lattice, the prescription
amounts to multiplying a U(1) phase factor to the gauge links.
Choosing $A_y = B (x-x_0)$, a constant magnetic field B can be introduced 
in the $z$-direction. Then the phase factor is in the y-links
\begin{equation}
U_y \rightarrow \exp{(iqa^2B (x-x_0))} U_y.
\end{equation}
Here $x_0$ is the origin of the phase factor which we choose to be the same as 
the quark source location.
In general, the computational demand of such calculations can be divided into three levels.
The first is a {\em fully-dynamical} calculation. For each value of external field, a new dynamical ensemble is needed that couples to u-quark (q=2e/3), d-and s-quark (q=-e/3). This requires a Monte Carlo algorithm that can treat the three flavors separately. Quark propagators are then computed on the ensembles with matching field values. This is very challenging and has not been attempted.
The second can be termed {\em re-weighting} in which a perturbative expansion of the action in terms of external field is performed.
The third can be called {\em U(1) quenched}: No field is applied in the Monte-Carlo generation of the gauge fields, only in the valence quark propagation in the given gauge background.
In this case, any gauge ensemble can be used to compute valence quark propagators. 

In this work, we use standard Wilson actions on $24^3\times 48$ lattice at $\beta=6.0$, both SU(3) and U(1) quenched, and  the following pion masses 796, 592, 563, 548, 533, 517 MeV.
The strange quark mass is set at the pion mass 592 MeV.  We analyzed 1000 configurations.
The point source location for the quark propagators is (t,x,y,z)=(0,12,12,12).
We considered 4 magnetic fields given in terms of the dimensionless number 
$\eta=qBa^2=0.001364\times n$ (for $n=1,2,3,4$) on the d-quark.
Of the 4 values, only the n=4 field satisfies the quantization condition for uniform magnetic flux 
in the xy plane which requires $\eta$ to be integer multiples of $2\pi/(N_xN_y)$\cite{rubin95}. 
To minimize boundary effects, we work with Dirichlet b.c. in the x and y directions  
and place the source in the center of the lattice.
We also apply Dirichlet b.c. in the time direction for longer time evolution. The  b.c. in the z direction is periodic.
To minimize possible contamination from higher power terms, we repeat the entire calculation 
with the magnetic field reversed. 
By taking the mass shift combination $(\delta m(B) - \delta m(-B))/2$, the even-powered terms 
are eliminated, so the contamination comes in at $O(B^3)$.
The added cost is further compensated by the fact that the same data set can yield information 
on the magnetic polarizability, 
by taking the average $(\delta m(B) + \delta m(-B))/2$ to 
eliminate the odd-powered terms in the mass shift~\cite{Lee05,Lee06}. 

The interaction energies for positive and negative-parity baryon states $E_\pm$ are extracted 
from the correlation function
\begin{eqnarray}
G(t) &=& \sum_{\vec{x}}\langle vac| \eta(x) \bar{\eta}(0)| vac \rangle \\
&=& 
(1+\gamma_t)\left[ A_+ e^{-E_+ (t-t_0)} + b A_- e^{-E_- (N_t+t-t_0)}\right]
+
(1-\gamma_t)\left[ b A_+ e^{-E_+ (N_t+t-t_0)} + A_- e^{-E_- (t-t_0)}\right]
\nonumber
\label{corr}
\end{eqnarray}
where $b=1, -1, 0$ correspond to periodic, anti-periodic, and Dirichlet boundary conditions in 
the t direction, respectively. In the case of the Dirichlet b.c. used in this work, 
parity projection is natural since $G(t)$ separates into two branches, one for each parity.
We use standard interpolating fields of the $C\gamma_5$ type for octet baryons.
For example, $\eta=\epsilon_{abc} \left( u^{aT} C\gamma_5 d^b \right) u^c $ for the proton.
The interpolating fields for other baryons can be found in Ref.~\cite{Lee06}.

 \section{Results and discussion}
\begin{figure}[tbh]
   \centering
   \includegraphics[width=8cm]{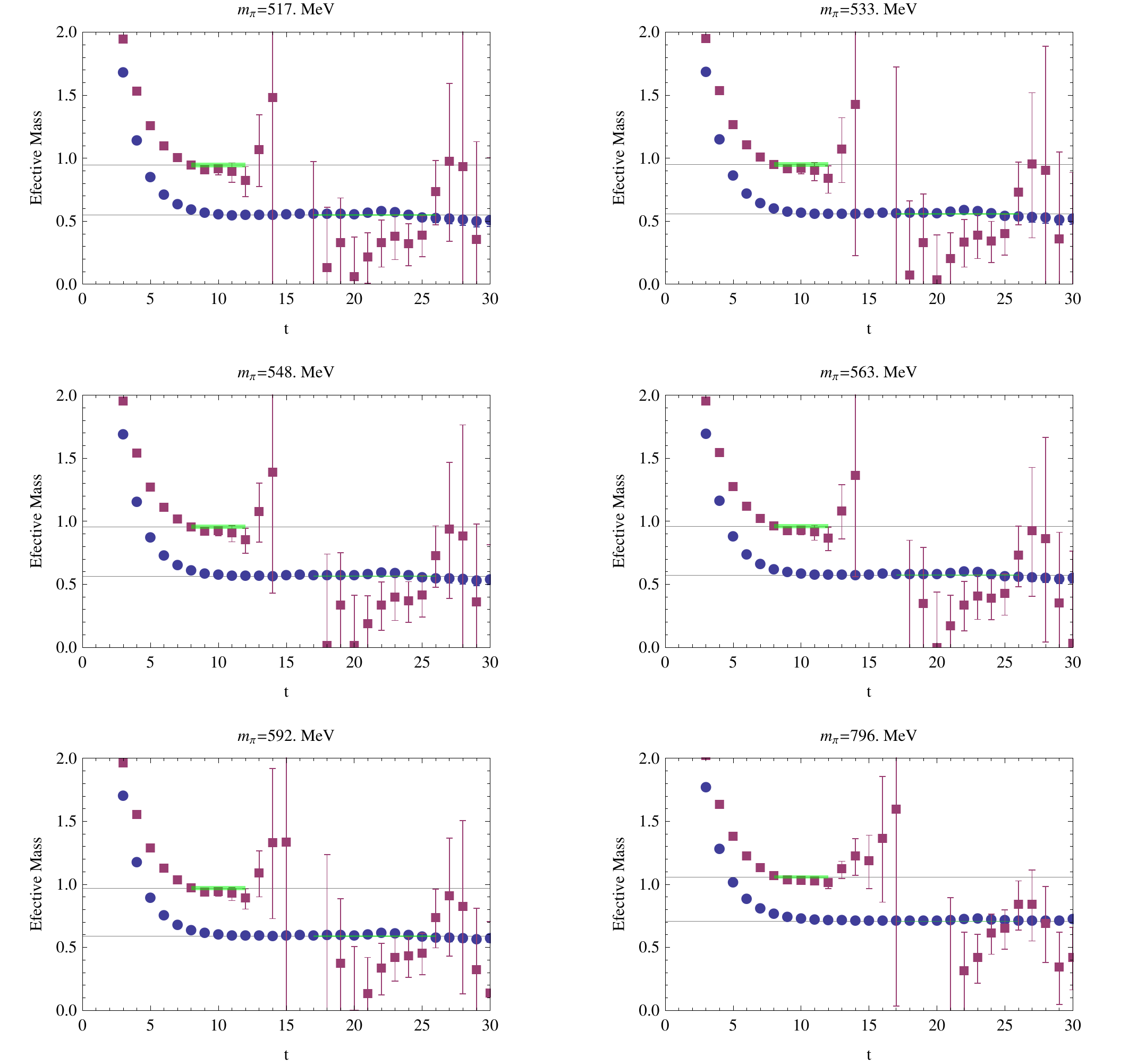} 
\hfill
   \includegraphics[width=7cm]{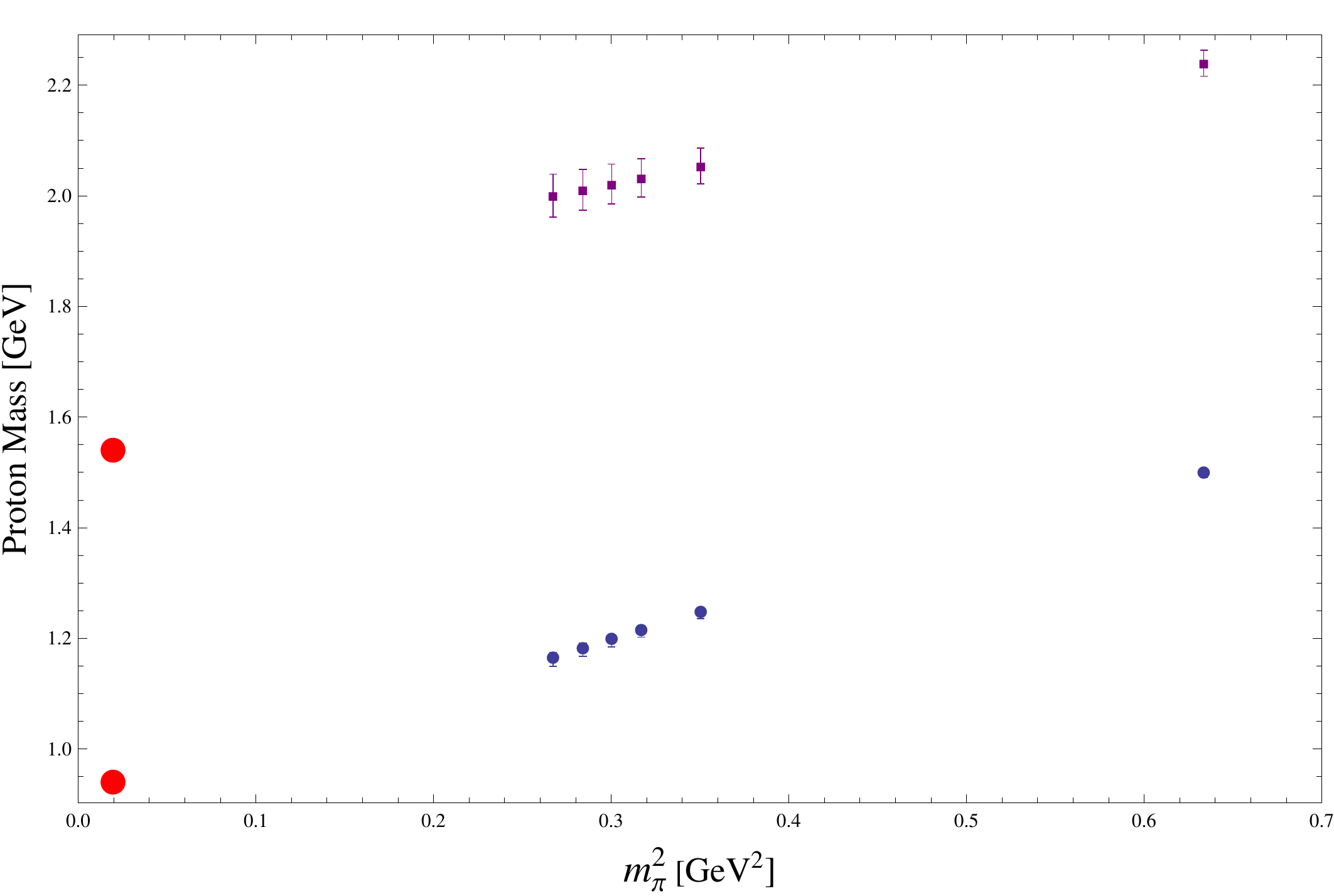}
   \caption{Effective mass plots (left) and the extracted masses (right) 
as a function of the pion mass squared at zero field in the nucleon channel. 
The (blue) circles are for positive parity, and the purple squares negative parity.}
   \label{emass}
\end{figure}

First we look at the signal for masses in the absence of magnetic fields.
Figure~\ref{emass} displays the effective mass plot in the nucleon channel
and the extracted masses for both parities. Very long plateaus are observed 
for positive parity, as expected. The signal for negative parity is much noisier and the 
plateau is much shorter. The fitted masses for positive parity in the time window of [17,26] 
and negative parity in [8,12] are also shown in this figure, along with experimental values. 
 
\begin{figure}[tbh]
   \centering
   \includegraphics[width=7cm]{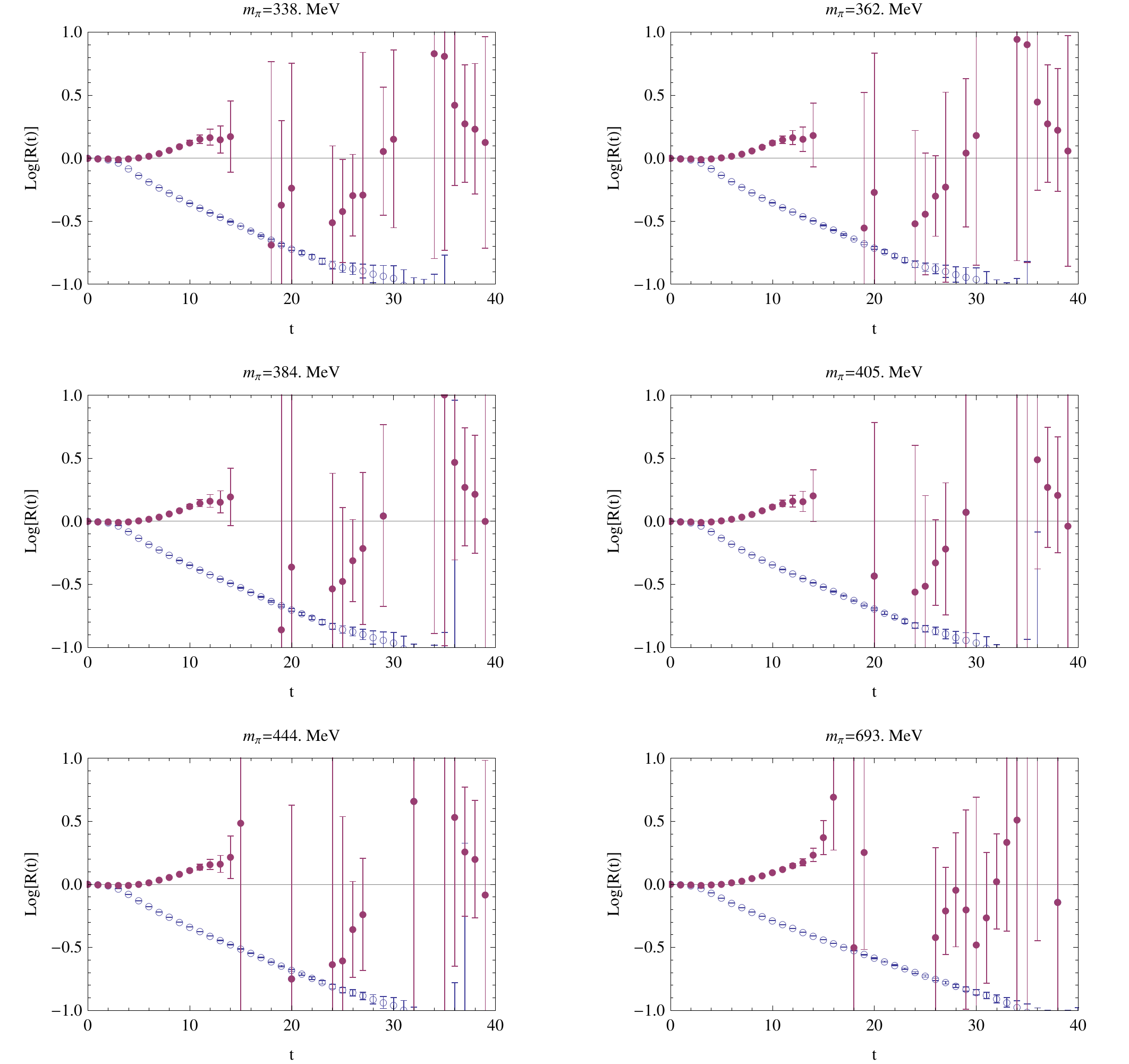}
\hfill
   \includegraphics[width=7cm]{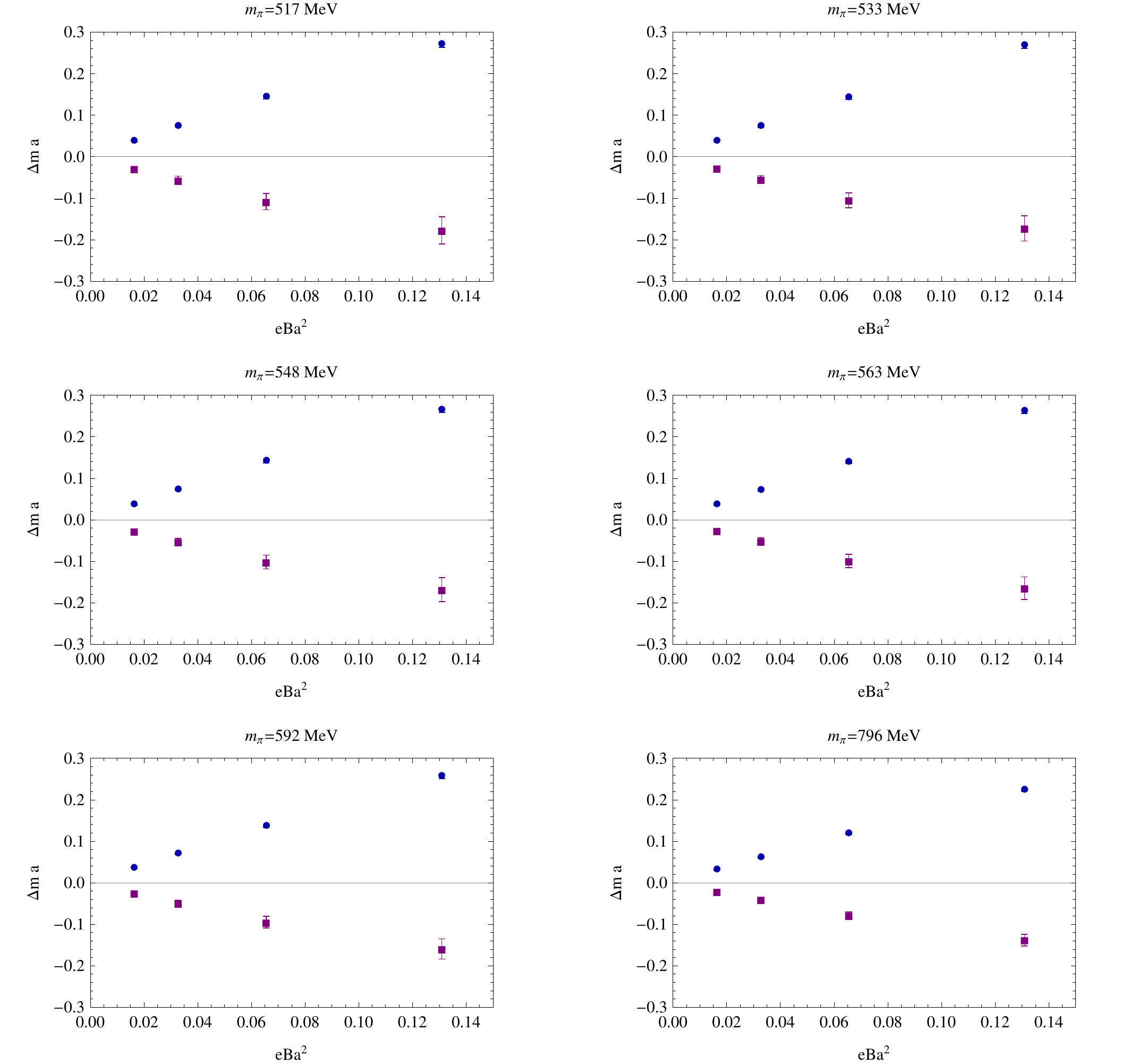}
   \caption{Left: the logarithm of the ratio defined in Eq.~(\protect\ref{ratio}) 
in the proton channel at the six pion masses for the smallest field.
Right: the extracted mass shifts as a function of the field at the six pion masses.}
   \label{proton-shift}
\end{figure}

To access the magnetic moments, we construct the following ratio of correlation functions
\begin{equation}
R(t,B)=\left[{G_+(t,B)\over G_-(t,B)}\right] /\left[ {G_+(t,-B)\over G_-(t,-B)}\right]
\label{ratio}
\end{equation}
where $G_+$ is the spin-up component of the correlation function and $G_-$ spin-down.
At large time, the ratio $R \rightarrow e^{-\Delta m t}$  where $\Delta m=4\mu B$.
This ratio would be unity in the absence of the field or for zero $\mu$.
For positive B values in $R(t,B)$, we expect positive/negative $R$ for  
negative/positive magnetic moment $\mu$, and the rate of change is controlled by 
the magnitude of $\mu$.
In Figure~\ref{proton-shift} we show the logarithm of the ratio for the proton channel 
at smallest positive field on the left. The result shows unambiguously that the sign of the 
negative-parity magnetic moment is opposite to that of its positive-parity partner.
On the right in the same figure are the extracted mass shifts at all four fields. The same fitting window of [17,26] and [8,12] were used to extract these mass shifts. 
Good linear behavior is observed as a function of the field, indicating that the fields chosen are indeed small, except at the strongest field where there is a slight deviation from 
linearity suggesting a small contamination from the $B^3$ term in the mass shift. 
For this reason, we do not use the data at the strongest field.

\begin{figure}[tbh]
   \centering
   \includegraphics[width=8cm]{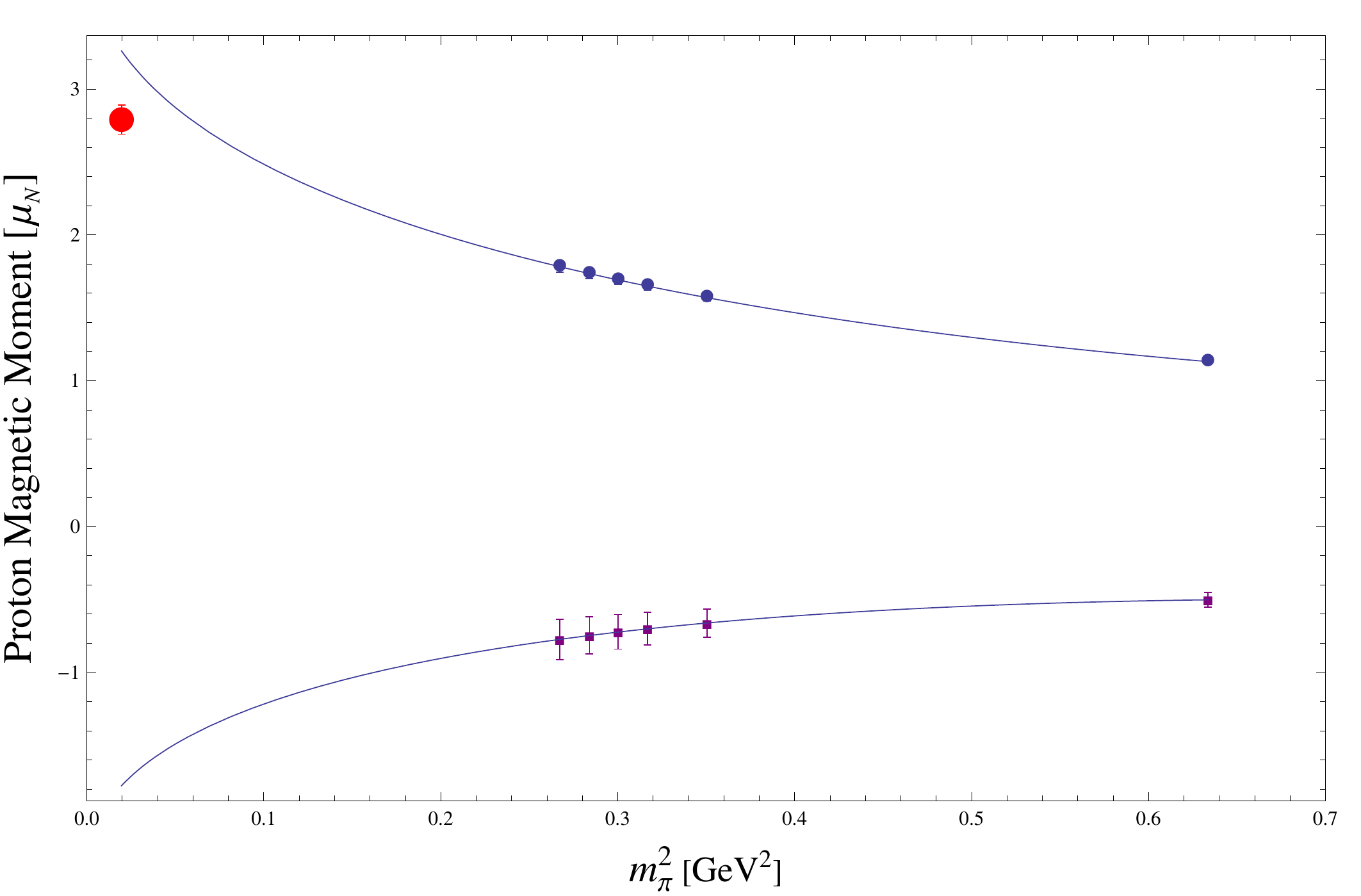}
   \caption{Magnetic moments in the proton channel.}
   \label{proton}
\end{figure}

Our results for the magnetic moments in the proton channel are displayed in Figure~\ref{proton}. 
There is some curvature in the data so we attempted a chiral extrapolation using the simple 
ansatz
\begin{equation}
\mu=a_0 + a_1 m_\pi+a_2 m_\pi^2,
\label{chiral2}
\end{equation}
where the $m_\pi$ term provides the non-analytic behavior in the quark mass.
Note that the magnetic moment defined in Eq.~(\ref{energy}) is in particle's natural 
magnetons.  To convert it into nuclear magnetons ($\mu_N$), we need to scale
the results by the factor $938/M$ where $M$ is the mass of the particle
measured in the same calculation at each pion mass.
We see that the extrapolation points to a result consistent with the experimental value 
of $2.79$ for positive parity, and predicts a negative value of $-1.8$ for negative parity.

\begin{figure}[tbh]
   \centering
   \includegraphics[width=8cm]{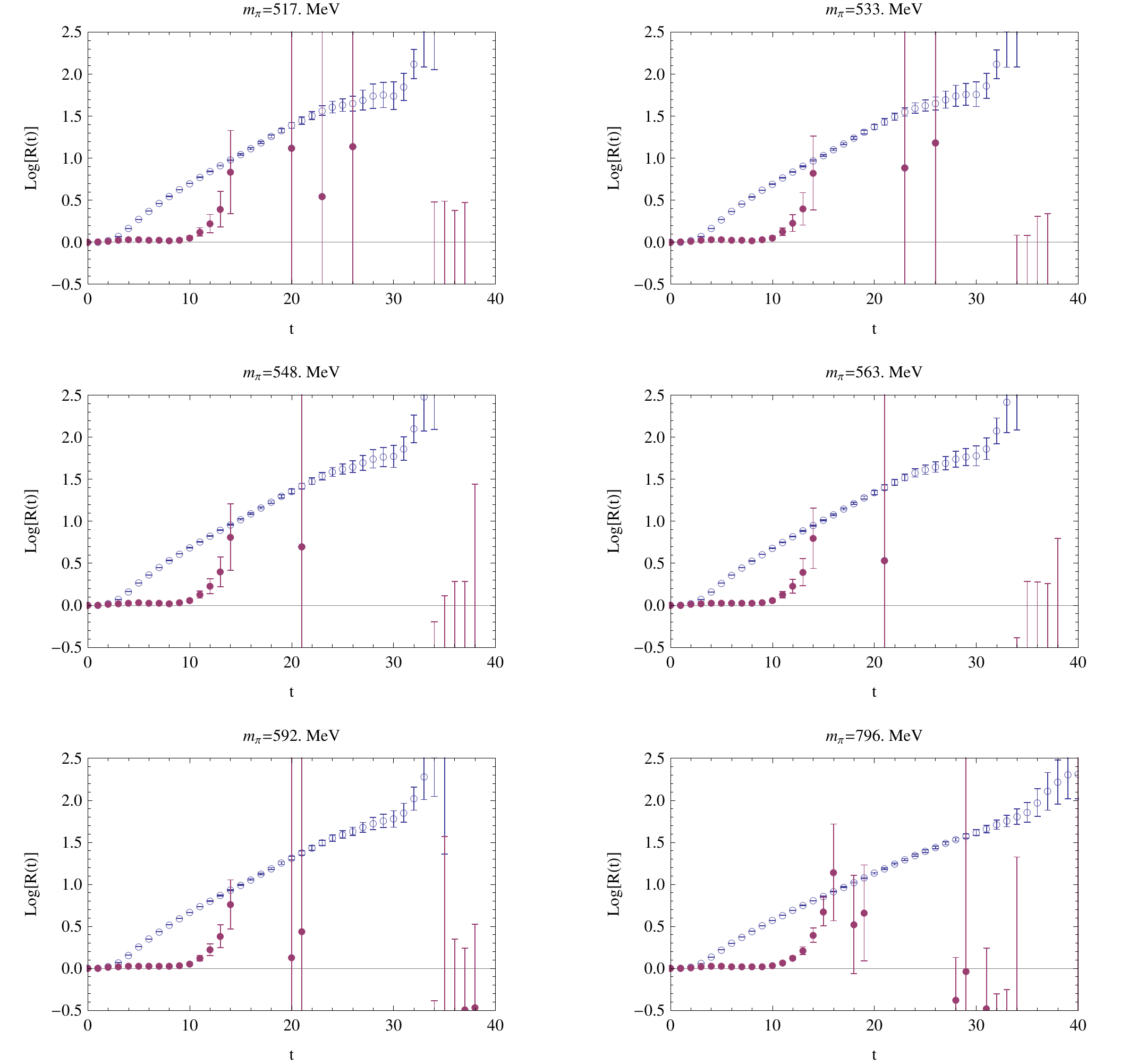} 
\hfill
   \includegraphics[width=7cm]{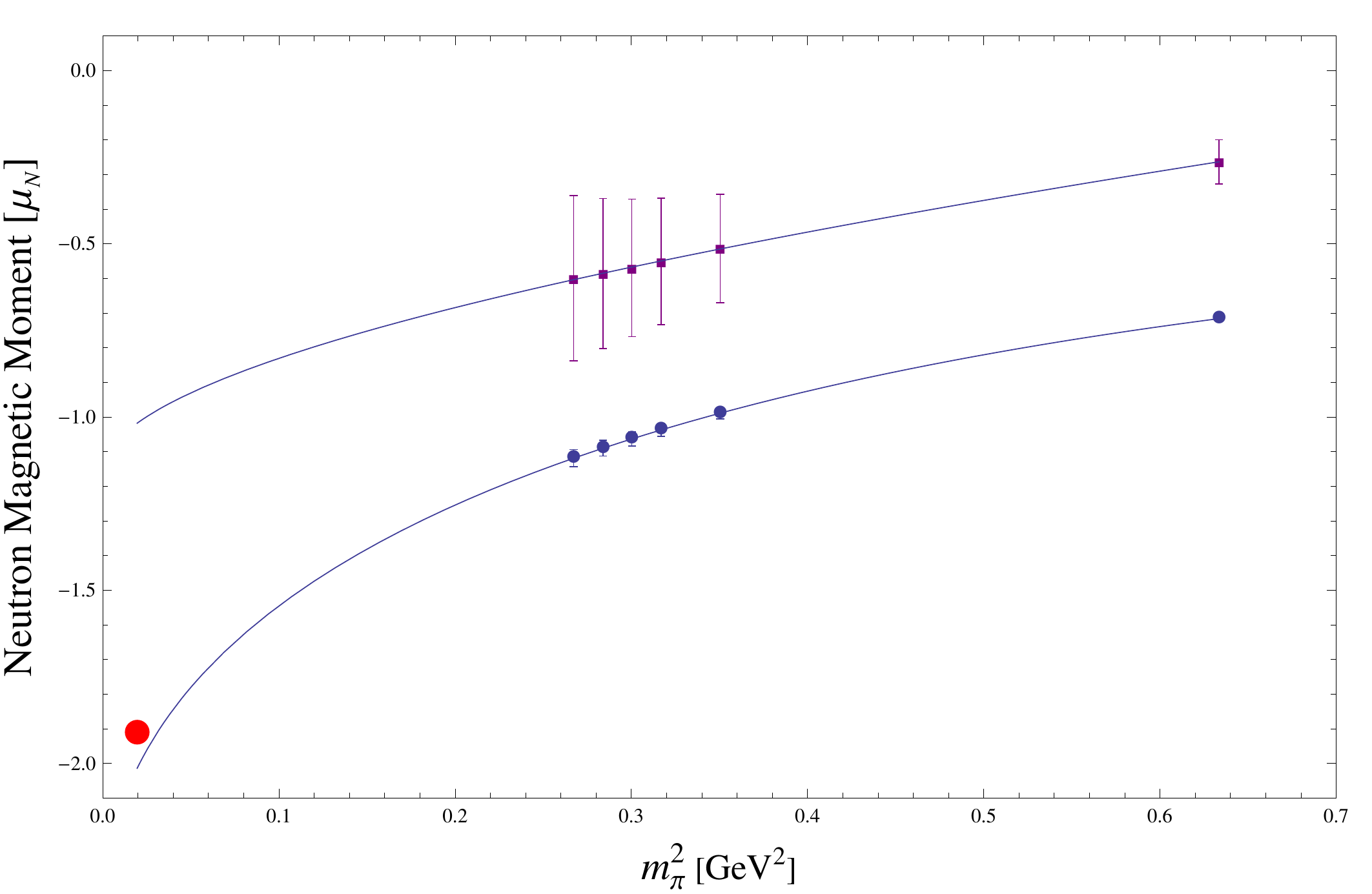}
   \caption{The same results in the neutron channel.}
   \label{neutron}
\end{figure}

Next we turn to the results in the neutron channel, as shown in Figure~\ref{neutron}.
Here we see that the magnetic moments have the same sign.
The extrapolated result for positive parity agrees with the experimental value of $-1.91$,
and predicts a negative value of about $-1.0$ for negative parity.
\begin{table}[htb] 
\caption{A comparison of magnetic moments for octet baryons.}
\label{comp}
\begin{tabular}{rccccc}
\hline\hline
State (spin-parity)  & Mass (MeV)& $\mu$ (Expt)& $\mu$ (LQCD)& $\mu$ (Unitary $\chi$PT)&$\mu$ (Quark Model)  \\
$p(1/2+)$            & 938       & 2.79  & 3.2         &              &      \\
$p^*(1/2-)$          & 1535      &       & -1.8        &  1.1         &  1.9 \\
\hline
$n(1/2+)$            &           & -1.91 & -2.0        &              &      \\
$n^*(1/2-)$          &           &       & -1.0        & -0.25        & -1.2 \\
\hline
$\Lambda_O(1/2+)$    & 1115      & -0.61 & -0.6        &              & -1.9 \\
$\Lambda^*_O(1/2-)$  & 1670      &       & -0.1        & -0.29        & -1.9 \\
\hline
$\Lambda_S(1/2-)$    & 1405      &       &  0          & 0.24 to 0.45 & 0.04 \\
$\Lambda^*_S(1/2+)$  &$\sim$2400 &       &  0          &              & -1.9 \\
\hline
$\Sigma^+(1/2+)$     & 1190      & 2.45  &  2.4        &              &      \\
$\Sigma^{+*}(1/2-)$  & 1620      &       &  -0.6       &              &      \\
\hline
$\Sigma^0(1/2+)$     &           &  0.65 &  0.8        &              &      \\
$\Sigma^{0*}(1/2-)$  &           &       &  0.1        &              &      \\
\hline
$\Sigma^-(1/2+)$     &           & -1.16 & -1.5        &              &      \\
$\Sigma^{-*}(1/2-)$  &           &       &  1.0        &              &      \\
\hline
$\Xi^0(1/2+)$        & 1320      & -1.25 & -1.0        &              &      \\
$\Xi^{0*}(1/2-)$     & 1690      &       & -0.5        &              &      \\
\hline
$\Xi^-(1/2+)$        &           & -0.65 & -0.5        &              &      \\
$\Xi^{-*}(1/2-)$     &           &       &  0.8        &              &      \\
\hline\hline
\end{tabular}
\end{table}

In fact, we surveyed all members of the baryon octet in order to establish a pattern.
In Table ~\ref{comp}, we summarize all of our results and compare them with 
experiment and other theoretical calculations.
The comparison is meant to be qualitative at this stage so no error bars are assigned. 
The statistical error on our results are less than 10\% for 
positive-parity states and 20\% for negative-parity states. We have not studied systematic errors.
In addition to the octet lambda ($\Lambda_O$), 
we also computed the flavor-singlet lambda ($\Lambda_S$).
The masses listed are standard values found in the particle data group,
except that of $\Lambda^*_S(1/2+)$ which we measured.
In fact, we measured the masses for all the channels 
(ignore isospin effects) at the six pion masses 
and used them to convert our magnetic moment results to nuclear magnetons. 
For the $1/2+$ states, it is encouraging to see that our results for the magnetic moments 
are consistent with experiment. 
Although not listed, we know the simple SU(6) quark model 
can largely reproduce this pattern, as well as a host of other models too numerous to cite here.
For the $1/2-$ magnetic moments, there exist limited theoretical studies based on simple quark model,
effective Lagrangian approach, and unitarized chiral perturbation theory~\cite{isobar02,Jido02,Jido03}.
It is interesting to see that our results disagree with model calculations by and large. 
For example, our result for $p^*(1/2-)$ has the opposite sign. 
For the $\Lambda_S$ channel, no signal for magnetic moments is observed in either parity.

 \section{Conclusion}
We have performed an exploratory study of the magnetic moments of spin-1/2, 
negative-parity baryons on the lattice using the background field method 
and standard lattice technology. 
The signal for $1/2+$ states is strong and robust and the results are consistent 
with experiment and other calculations.
Against that backdrop, the signal for $1/2-$ states is more limited but nonetheless discernible.
Relatively high statistics (over 800 configurations) are required to stabilize the signal.
A preliminary pattern across the entire spectrum is revealed for the first time.
The most intriguing result is that the sign of the  $p^*(1/2-)$ state is opposite 
to that expected from other theoretical calculations.
An experimental measurement in the nucleon channel (both proton and neutron), 
as proposed in Refs.~\cite{isobar02,Jido02,Jido03}, would be interesting.
Overall, our results demonstrate that the methodology is robust and relatively inexpensive
(only mass shifts are required). 

The calculation can be improved in a number of ways. 
The most important is the better isolation of the signal for the negative-parity states. 
This could be achieved by using smeared sources and/or a finer resolution in the time evolution, coupled with high statistics.
A study of systematic errors such as chiral extrapolations and finite-volume effects is 
also in order.

\section*{Acknowledgment}
This work is supported in part by U.S. Department of Energy
under grant DE-FG02-95ER40907. 
The computing resources at NERSC and JLab have been used.


\begin{thebibliography}{99}

\bibitem{Mahbub10} M.S. Mahbub, W. Kamleh, D.B. Leinweber, A.O. Cais, A.G. Williams,
Phys. Lett. {\bf B693}, 351 (2010).

\bibitem{Bulava10} J. Bulava, R.G. Edwards, E. Engelson, B. Joo, H-W. Lin, C. Morningstar, D.G. Richards, S.J. Wallace,
Phys. Rev. {\bf D92}, 014507 (2010).

\bibitem{Mathur05} N. Mathur, Y. Chen, S.J. Dong, T. Draper, I. Horvath, F.X. Lee, K.F. Liu, J.B. Zhang,
Phys. Lett. {\bf B605}, 137 (2005).

\bibitem{Sasaki02} S. Sasaki, T. Blum, S. Ohta,
Phys. Rev. {\bf D65}, 074503 (2002).




\bibitem{mart82} G. Martinelli {\it et al.},
Phys. Lett. {\bf B116}, 434 (1982).

\bibitem{bernard82} C. Bernard, T. Draper, and K. Olynyk,
Phys. Rev. Lett. {\bf 49}, 1076 (1982).

\bibitem{rubin95} H.R. Rubinstein, S. Solomon, and T. Wittlich,
Nucl. Phys.  {\bf B457}, 577 (1995).

\bibitem{Lee05} F.X. Lee, R.Kelly, L. Zhou, and W. Wilcox,
Phys. Lett. {\bf B627}, 71 (2005).

\bibitem{Lee06} F.X. Lee, L. Zhou, W. Wilcox, and J. Christensen, 
Phys. Rev. {\bf D73}, 034503 (2006).

\bibitem{Engel07} M. Engelhardt,
Phys. Rev. {\bf D76}, 114502 (2007).

\bibitem{Aubin09} C. Aubin, K. Orginos, V. Pascalutsa, M. Vanderhaeghen,
Phys. Rev. {\bf D79}, 051502 (2009).

\bibitem{Detmold10} W. Detmold, B.C. Tiburzi, A. Walker-Loud,
Phys. Rev. {\bf D81}, 054502 (2010).  


\bibitem{isobar02} W.H. Chiang and S.N. Yang,
arXiv:nucl-th/0211061.

\bibitem{Jido02} D. Jido, A. Hosaka, J.C. Nacher, E. Oset, and A. Ramos, 
Phys. Rev. C {\bf 66}, 025203 (2002).

\bibitem{Jido03} T. Hyodo, S.I. Nam, D. Jido, and, A. Hosaka,
arXiv:nucl-th/0305023.

\end{thebibliography}
\end{document}